\documentclass[prm,twocolumn, amsmath,amssymb,
reprint,%
author-year,%
author-numerical,%
dvipdfmx,
]{revtex4-2}

\usepackage{graphicx}
\usepackage{bm}
\usepackage{color}

\begin{document}

\title{Beyond geometrical screening in predicting two-dimensional materials}
\author{Shota Ono}
\email{shotaono@muroran-it.ac.jp}
\affiliation{Department of Sciences and Informatics, Muroran Institute of Technology, Muroran 050-8585, Japan}

\begin{abstract}
This perspective overviews the family of two-dimensional (2D) materials, which have attracted significant attention due to their properties and potential applications, and discusses how novel 2D materials including van der Waals (vdW) and non-vdW 2D materials have been predicted so far. A few thousand 2D materials have been predicted to be exfoliable or dynamically/thermodynamically stable, whereas a few hundred 2D materials have been synthesized so far, highlighting a gap between the theoretical prediction and experiments. This perspective introduces the recent developments in predicting the synthesis of non-vdW 2D materials. 
\end{abstract}

\maketitle


\section{Introduction}
Since the exfoliation of monolayer graphene from graphite, two-dimensional (2D) materials have attracted much attention due to their physical and chemical properties that are markedly different from those of 3D counterparts. Graphene has a linear dispersion at the K point around the Fermi level, providing a platform for exploring the Dirac fermion physics, and its unique properties enable numerous applications \cite{Novoselov2012}. 2D transition metal dichalcogenides $MX_2$, where $M$ is the transition metal element and $X$ is the chalcogen atom, adopt several phases such as 2H, 1T, and 1T$^\prime$, governing the electronic, magnetic, and mechanical properties. Thus, phase engineering by strain and doping as well as the combination of $M$ and $X$ has been extensively studied, leading to next-generation electronics and valleytronics \cite{Zhao2021}.   

Beyond graphene, several 2D elemental materials termed Xenes have been explored. Most of them are classified as non-van der Waals (non-vdW) 2D materials because their 3D counterparts have non-layered structure. First-principles calculations have predicted that silicene (Si) and germanene (Ge) have buckled honeycomb structure that is different from the planar geometry of graphene \cite{Cahangirov2009}. Following the theoretical prediction, the experimental synthesis of silicene \cite{Vogt2012,Fleurence2012} and germanene \cite{Li2014,Bampoulis_2014,Davila_2014} were reported by several groups. Realization of stanene (Sn) \cite{Zhu2015} and plumbene (Pb) \cite{Yuhara2019} were also reported. Borophene (B) was cleaved from $\beta$-rhombohedral boron \cite{Chung2024}. The stability of 2D metals has been systematically investigated using first-principles approach \cite{Nevalaita2018,SO2020}. More recently, goldene has been synthesized, exhibiting a hexagonal monolayer structure derived from the Au(111) surface \cite{Kashiwaya2024}. A recent review discusses the details of experimental synthesis routes for Xenes \cite{Molle2025}. The physical properties of non-vdW 2D materials have been investigated, such as quantum spin Hall states in group IV 2D materials \cite{Acun_2015,Bampoulis2023} and in-plane transport properties of goldene \cite{Zhao2024}. 

\begin{figure}
\center\includegraphics[scale=0.4]{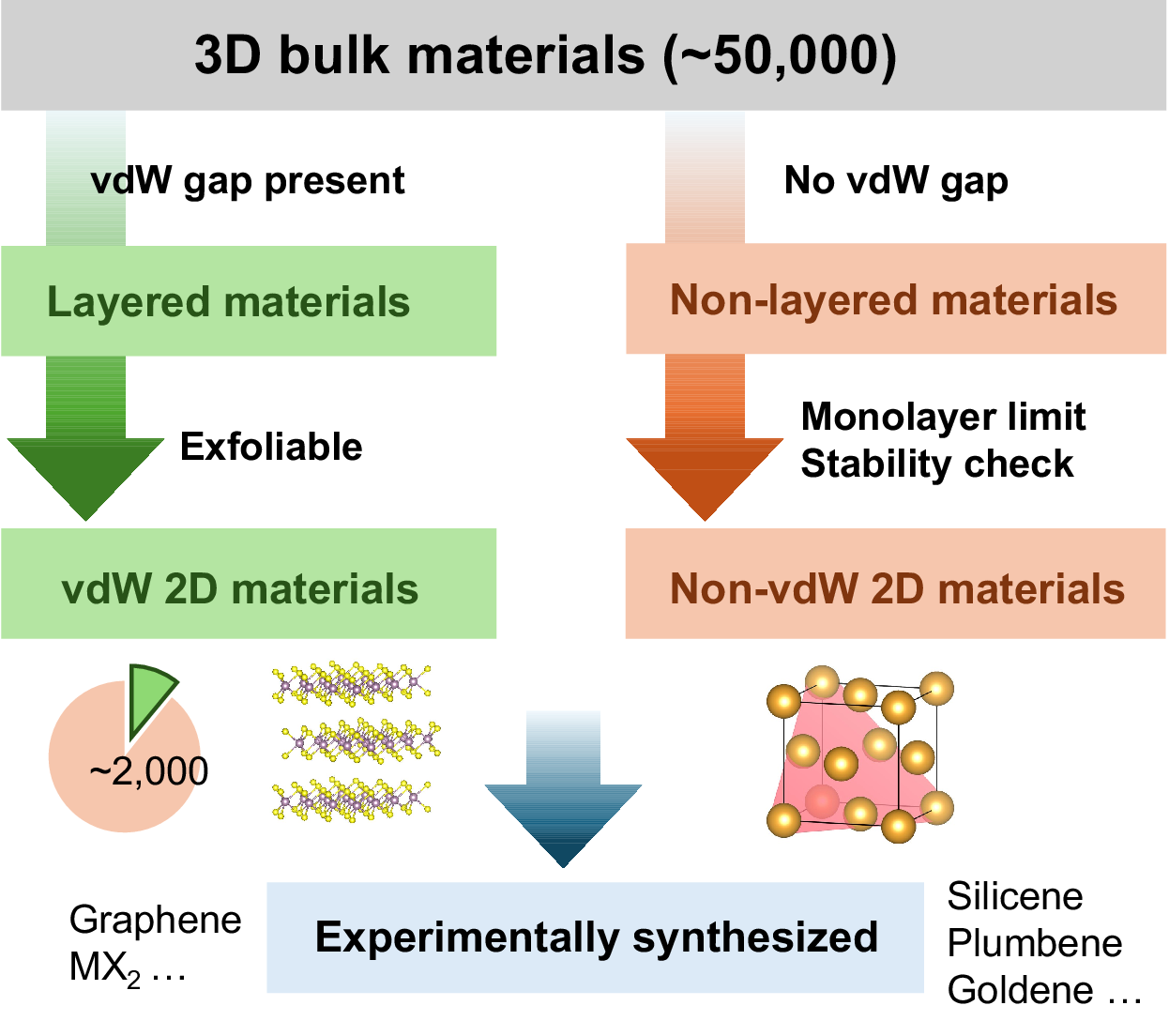}
\caption{Relationship between 3D and 2D materials. Through high-throughput screening of about 50,000 bulk materials, a few thousand potential vdW 2D materials are predicted to be exfoliable from layered materials. On the other hand, non-vdW 2D materials are identified by studying the stringent stability check for their monolayer structures. A few hundred 2D materials are synthesized experimentally. The crystal structure is visualized by using VESTA \cite{vesta}. } \label{fig1} 
\end{figure}

The family of 2D materials has rapidly expanded in recent years, driven by the development of computational materials databases constructed from high-throughput first-principles calculations with the help of machine-learning approaches. Several thousand candidate 2D materials have been predicted \cite{Lebegue2013,Ashton2017,Mounet2018,Haastrup_2018,Gjerding_2021}, and several databases have been used to explore their functional properties. However, only a few hundred 2D materials have been experimentally synthesized (see X2DB \cite{X2DB}), highlighting limitations in current screening strategies. In particular, reliable approaches for predicting non-vdW 2D materials have not been established, as the stabilization mechanisms are fundamentally different from those of vdW 2D materials (see Fig.~\ref{fig1} for the classification of 2D materials). Therefore, it is timely to reexamine the construction principle underlying existing 2D materials databases. 

In this perspective, we review how 2D materials have been explored and predicted through screening materials space. This will help us to understand which 2D materials are actually synthesized experimentally, which is necessary for any practical applications. Therefore, we focus on structural stabilities in the following sections.    

\section{How to predict 2D materials}
The Inorganic crystal structure database (ICSD), the world's largest materials database, contains more than 327,833 materials (on October 21, 2025). The Materials Project (MP) contains about 154,879 materials, among which 49,283 materials have an ICSD number (on September 25, 2025) \cite{MP2013}. Several approaches have been proposed to estimate the number of possible 2D materials that can be created from 3D structures.  

Before considering the approaches used in the prediction of 2D materials, it should be emphasized how the materials prediction has been done in the field of computational materials science. The ``stable material'' should have (i) non-positive formation energy to avoid the decomposition into different constituents and (ii) positive vibrational frequencies over the entire Brillouin zone to avoid the phase transition into different crystal structures at zero and finite temperatures \cite{Malyi2019}. In general, the dynamical stability at zero and finite temperatures are investigated through phonon dispersion calculations and molecular dynamics simulations, respectively. To accurately predict the quadratic dispersion of out-of-plane vibrational modes, it is important to consider the invariance and equilibrium conditions for interatomic force constants \cite{lin_general_2022}. If a material satisfies the conditions (i) and (ii), such a material is regarded as stable. However, this does not guarantee the experimental realization of the predicted material: although the two conditions may allow the existence of the local minimum in the potential energy surface (PES) characterized by many degrees of freedom (lattice parameters, atomic positions, and atomic species), their conditions do not provide how to reach the energy minimum. It is desirable to include kinetic effects and growth pathways of the materials synthesis in the computational simulation, which remains an open question in this research field. 

\subsection{Geometrical screening}
Graphene and $MX_2$ are classified into vdW 2D materials, where the interlayer bonds are much weaker than the intralayer bonds in the parent 3D crystal. The vdW 2D materials not only exhibit interesting properties themselves, but also open up the Moir{\'e} physics in multilayered heterostructures. Exploring novel vdW 2D materials is crucial to advance the 2D materials physics. 

We first discuss a top-down approach that is applicable to the prediction of vdW 2D materials. Leb\`{e}gue et al. have focused on the geometry of the crystal structures contained in 3D materials database \cite{Lebegue2013}. They identified layered structures by investigating whether the packing ratio is small (0.15--0.5) and the interlayer distance along the $c$ axis is sufficiently large ($>2.4$ \AA). This screening approach was improved by several research groups \cite{Ashton2017,Larsen2019,Mounet2018,Haastrup_2018,Gjerding_2021}. For example, a vdW gap is identified if the interatomic distance $d_{ij}$ between atoms $i$ and $j$ satisfies the inequality 
\begin{eqnarray}
d_{ij} > k (r_i + r_j),
\end{eqnarray}
where $r_i$ is the atomic radius of atom $i$ and $k$ is a parameter \cite{Larsen2019}. On the other hand, $d_{ij} < k (r_i + r_j)$ indicates the presence of covalent bond between atoms $i$ and $j$. In this way, the chemical bonds are identified for each 3D material. A different inequality such as $d_{ij} > r_i + r_j + \Delta$ has also been used, where $\Delta$ is a parameter \cite{Mounet2018}. Once different atoms are bonded in the unit cell, a topology-scaling approach (TSA) developed by Ashton et al. was used to determine the material's dimensionality \cite{Ashton2017}. In TSA, one identifies the cluster in the unit cell and also in an $n\times n\times n$ supercell. If the number of atoms in the cluster scales as $n^3$ and $n^2$, then the crystal is classified into 3D and 2D, respectively. Although this is useful for the dimension classification, it remains unclear how to determine appropriate values of $k$. Larsen et al. introduced a score for determining the material's dimensionality \cite{Larsen2019}. They studied how the dimensionality changes by varying the value of $k$ and determined a relevant dimensionality within the range of $k$. This scheme is implemented to Atomic Simulation Environment (ASE) \cite{ASE}. In this way, geometrical screening of a materials database identifies possible layered structures.  

Given layered materials, we next study how easy the monolayer is exfoliated from the bulk. This is investigated by calculating the binding energy per area, $E_{\rm b} = (E_{1} - E_{N}/N)/A$, where $E_{n}$ is the total energy of $n$ layer thin films and $A$ is the in-plane area of the bulk unit cell. In the thermodynamic limit $N\rightarrow \infty$, $E_N/N$ becomes the total energy of bulk $E_{\rm bulk}$. Using this relation ($E_N/N \rightarrow E_{\rm bulk}$), Jung et al. have shown that the binding energy, $\lim_{N\rightarrow \infty} (E_{1} - E_{N}/N)$, is equivalent to the exfoliation energy, $\lim_{N\rightarrow \infty} (E_1 + E_{N-1} - E_N)$ \cite{Jung2018}. Based on this theorem, the exfoliation energy per area is defined as
\begin{eqnarray}
 E_{\rm exf} = \frac{E_{1} - E_{\rm bulk}}{A}.
\end{eqnarray}
Graphene and MoS$_2$ exhibit exfoliation energies of 21 and 18 meV/\AA$^2$, respectively. These values are used to identify novel 2D materials that are easily exfoliated from the layered structures.  

Through geometrical screening of materials database, Leb\`{e}gue et al. have identified 92 potential 2D materials \cite{Lebegue2013}. In fact, ferromagnetic 2D VSe$_2$ was synthesized experimentally \cite{bonilla_2018}. By screening the MP database, Ashton et al. have identified 826 layered materials using TSA \cite{Ashton2017}. Among them, 680 2D materials are potentially exfoliable through first-principles calculations of the exfoliation energy. Mounet et al. have identified 1,825 2D materials by extensively screening the ICSD and COD \cite{Mounet2018}. This has been expanded to 3,077 in total by screening an additional database (MPDS) and the updated version of the two databases \cite{Mounet2023}. 

In this way, a few thousand 2D materials have been identified. However, only a few hundred 2D materials have been experimentally realized \cite{X2DB}. The discrepancy between the theoretical prediction and experiments may be attributed to (i) the deficiency of essential factors characterizing the stability of 2D materials in the computational screening processes and (ii) no experimental trials to create the predicted 2D materials. It is important to fill the gap in 2D materials science. 

\subsection{Beyond geometrical screening}
The geometrical screening combined with the exfoliation energy calculations enables us to identify the layered structures that can yield vdW 2D materials. However, this scheme is not applicable to identify non-vdW 2D materials such as silicene, germanene, and goldene because no vdW gap exists in 3D bulk. Also, 2D compounds without 3D counterparts are excluded {\it ab initio}. Many potential 2D materials might be overlooked in the geometrical screening of materials database.

To go beyond the top-down approach, Haastrup et al. have constructed a work flow for combinatorial lattice decoration \cite{Haastrup_2018}. They started from more than 40 structure prototypes (e.g., MoS$_2$- and CdI$_2$-type structures for $AB_2$) and created possible 2D materials by lattice decoration (e.g., $A$ for transition and noble metal elements and $B$ for chalcogen and halogen atoms in $AB_2$). The thermodynamic and dynamical stabilities have been investigated by calculating the formation energy and the phonon dispersions, respectively. Computational 2D materials database (C2DB) includes more than 4,000 stable structures and their physical properties such as electronic band structure, elastic constants, and magnetic properties \cite{Gjerding_2021}. The number of prototypes has been extended using machine learning approaches \cite{lyngby_2022,Wang_2023}. 

The lattice decoration approach has been used for the family of $MA_2Z_4$ ($M=$ transition metals; $A=$ Si and Ge; $Z=$ N, P, and As). MoSi$_2$N$_4$ is an artificially created vdW 2D material because no 3D counterpart exists \cite{Hong2020}. The MoS$_2$-type structured MoN$_2$ monolayer is sandwiched with the outer N-Si layers, where seven atomic layers are ordered as N-Si-N-Mo-N-Si-N. Through lattice decoration for $MA_2Z_4$, several 2D systems have been predicted, and these systems cover various electronic phases such as superconducting and ferromagnetic phases \cite{wang_intercalated_2021}. Thus far, the syntheses of MoSi$_2$N$_4$ and WSi$_2$N$_4$ are reported \cite{Hong2020}. 

The geometrical screening searches for anisotropy of crystal structures in real space. However, this is not sufficient. Jia et al. have focused on the anisotropy of elasticity in 3D crystals \cite{Jia2021}. They assumed that there are four types of crystals including exfoliable layered crystal, non-exfoliable layered crystal, non-exfoliable non-layered crystal, and exfoliable non-layered crystal. They pointed out that the exfoliable non-layered crystal are wrongly excluded from the candidate 2D materials in the geometrical screening. To resolve this issue, they introduced a measure of exfoliability defined as
\begin{eqnarray}
 E = \max_{\bm{z}} \left[ \frac{{\rm min}_{xy} (Y_{\rm in})}{Y_{\rm out}}\right],
\end{eqnarray}
where $Y_{\rm in}$ and $Y_{\rm out}$ are the in-plane and out-of-plane Young's modulus, and $\bm{z}$ is the out-of-plane directional vector of an arbitrarily possible exfoliable plane $xy$. Maximizing the ratio of Young's modulus corresponds to maximize the structural anisotropy of materials. They analyzed 10,812 crystals in MP database and identified 148 exfoliable crystals. Among them, 34 predicted crystals have been experimentally exfoliated into 2D materials. They have found LiC$_6$ and Ca$_2$Cu(ClO)$_2$ as exfoliable non-layered crystals, which wait to be synthesized experimentally. 

The elastic constants are macroscopic quantities of solids. To study the anisotropy in a microscopic manner, Bagheri et al. have focused on the force constants (FCs) that reflect the bond strength between different atoms \cite{Bagheri2023}. They defined the maximum and minimum value of FC as
 \begin{eqnarray}
 {\rm MaxFC} &=& \max_{i} \left[ \max_{j\ne i} \left( \left\Vert \Phi_{ij}^{\alpha\beta} \right\Vert_{\rm F}\right) \right], \\
 {\rm MinFC} &=& \min_{i} \left[ \max_{j\ne i} \left( \left\Vert \Phi_{ij}^{\alpha\beta} \right\Vert_{\rm F}\right) \right], 
 \end{eqnarray}
where $\Phi_{ij}^{\alpha\beta} = \partial^2 U/\partial r^{\alpha}_{i}\partial r^{\beta}_{j}$ is the force constant matrix, the second derivative of the potential energy $U$ with respect to the $\alpha$-component of the $i$th atom position $r_{i}^{\alpha}$. After taking the Frobenius norm $\Vert \Phi_{ij}^{\alpha\beta} \Vert_{\rm F} \equiv \Phi_{ij}$, the maximum FC, $\Phi_{i}^{\rm max} = \max_{j\ne i}(\Phi_{ij})$, is calculated for each atom $i$. Therefore, MaxFC and MinFC reflect the strongest and weakest bonds between atoms in the unit cell. They analyzed 10,032 materials obtained from the phonon database developed by Togo \cite{TOGO20151}. In addition to various vdW 2D materials, they have found several oxides to be exfoliable non-layered 2D materials. Recently, this search has been extended using universal machine-learning interatomic potentials \cite{Bagheri2026}.

\subsection{Non-van der Waals 2D materials}
High-throughput screening of 3D materials database is useful to identify potential 2D materials. However, without studying the  geometry relaxation of 2D structures in detail, high-throughput screening may overlook many 2D materials hidden in 3D structures. For example, let us assume that ultrathin films are truncated from non-layered (non-vdW) materials. Such thin films  inevitably have dangling bonds. Then, they will exhibit strong surface modification and/or structural transformation, producing a significant energy gain and yielding non-vdW 2D materials. In this section, several approaches exploring such 2D materials are introduced. 

\subsubsection{Prototype-based search}
We first consider traditional semiconductors in the monolayer limit. The chemical bond is an important concept in classifying traditional semiconductors: the covalent bond character decreases, whereas the ionic character increases, as one goes from III-V to II-VI to I-VII semiconductors. The III-V semiconductors have the zincblende-type structure, whereas II-VI and I-VII semiconductors have the rocksalt-type or cesium chloride-type structures. The stable 2D structures of these semiconductors have been explored \cite{Zheng_2015,Lucking2018,SO2022}. For example, III-V semiconductors exhibit the double layer honeycomb structure in the monolayer limit \cite{Lucking2018}. Such a structure is derived from the (111) surface, accompanying in-plane atomic displacements to enhance the bonding strength between upper and lower layers. Following the theoretical prediction, the experimental synthesis of 2D AlSb has been reported \cite{Qin_2021}. This highlights the importance of the geometry relaxation in the monolayer limit. 

Atomically thin minerals, such as hematite (Fe$_2$O$_3$) and ilmenite (FeTiO$_3$), have been synthesized \cite{Balan_2018_hematite,Balan_2018_ilmenite}. Motivated by these syntheses, Friedrich et al. have systematically studied non-vdW 2D materials by focusing on the binary and ternary compounds with the same structures as hematite and ilmenite \cite{Friedrich2022}. They have screened the AFLOW-ICSD database and found 8 binary and 20 ternary oxides. They calculated the exfoliation energy $E_{\rm exf}$ of 2D materials derived from the (0001) surface of 3D bulk. The hematene and ilumenene have $E_{\rm exf}\simeq 140$ and 100 meV/\AA$^2$ that are larger than 20 meV/\AA$^2$ of graphene. They proposed that the surface should be terminated by elements with low oxidation numbers (e.g., Na, Cu, and Ag in state $+1$) to reduce the value of $E_{\rm exf}$. 

\begin{figure*}
\center\includegraphics[scale=0.5]{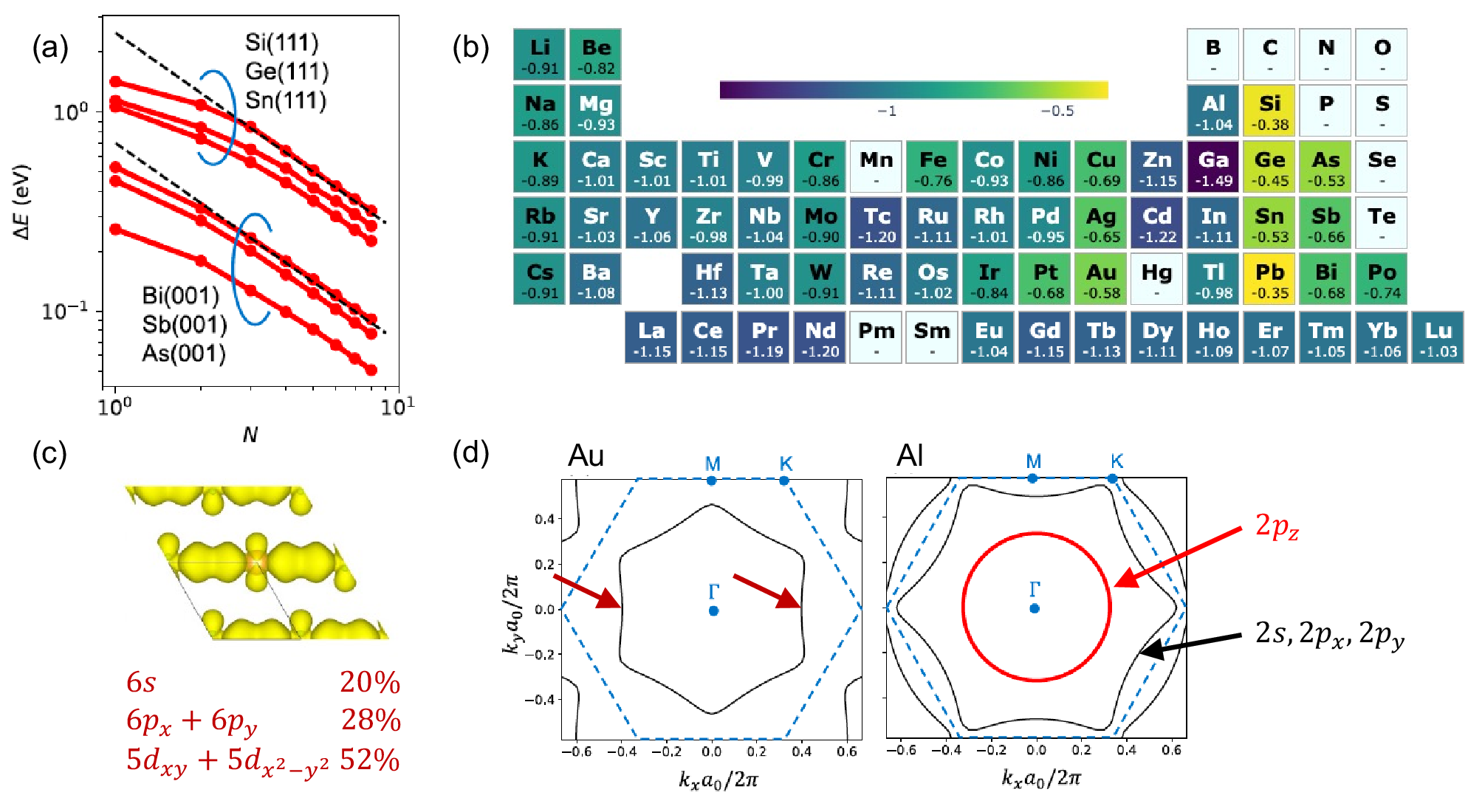}
\caption{(a) The $N$-dependence of FTEE defined by Eq.~(\ref{eq:FTEE}). In the monolayer limit, group 14 and 15 elements exhibit a deviation from the $N^{-1}$ law, indicating that those thin films enjoy 2D structures. (b) The values of $p(1)$ defined by Eq.~(\ref{eq:pN}) for elements in the periodic table. The value of $p(2)$ is indicated for Pb because the FTEE is anomalously small at $N=2$, corresponding to the plumbene in the buckled honeycomb structure. (c) The electron charge density at the Fermi surface edge, indicated by arrows in (d) for goldene. (d) Fermi surface of goldene and aluminene. The calculations were performed using Quantum Espresso \cite{qe}. Data reproduced from Ref.~\cite{SO2025,SO_HY_2025}. } \label{fig2} 
\end{figure*}

\subsubsection{3D-2D transition-based search}
Many non-vdW 2D materials have been predicted through the cohesive, formation, and exfoliation energy calculations, phonon dispersion calculations, and molecular dynamics simulations. These results provide a necessary condition for the stable 2D structure. However, the underlying physics behind the creation of non-vdW 2D materials should be attributed to how the material prefers 2D structures when reducing the thickness of their ultrathin films, rather than whether the 2D structure is located at the local minimum in the PES. Below, the author's work considering this issue is introduced \cite{SO2025}. 

To study how a parent 3D material adopts 2D structures, let us define the finite-thickness excess energy (FTEE) as 
\begin{eqnarray}
\Delta E_\alpha (N) = \frac{E_\alpha (N)}{N}-E_{\rm bulk},
\label{eq:FTEE}
\end{eqnarray}
where $E_\alpha (N)$ and $E_{\rm bulk}$ are the total energy per unit cell of $N$-layer thin film and bulk, respectively. The subscript $\alpha$ is the index specifying the surface cut. The FTEE includes the excess energy due to the surface and finite-size effects, which becomes negligible in the bulk limit. We can show the following relation for large $N$: 
\begin{eqnarray}
 \Delta E_\alpha \propto N^{-1}.
 \label{eq:3D}
 \end{eqnarray}
To prove this, assume that the surface relaxation occurs only at the top and bottom layers. Then, the $N$ layer thin film consists of $(N-2)$ bulk layers and two surface layers. For large $N$, one can ignore the interaction between the upper and bottom surfaces, and express the total energy as $E_\alpha(N) \simeq (N-2)E_{\rm bulk}+2E_{\rm surf}^{\alpha}$, where $E_{\rm surf}^{\alpha}$ is the energy of the $\alpha$ surface layer. Thus, one obtains $\Delta E_\alpha (N) = E_0/N$, where the factor $E_0=2(E_{\rm surf}^{\alpha} - E_{\rm bulk})$ is independent of $N$. 

It should be emphasized that if Eq.~(\ref{eq:3D}) holds, such a thin film may be regarded as a 3D-like material. On the contrary, if a breakdown of Eq.~(\ref{eq:3D}) is observed, such a thin film is no longer 3D-like. Furthermore, if a downward deviation from the $N^{-1}$ scaling is observed in the monolayer limit, this should indicate a 3D-2D transition, i.e., non-vdW 2D material is stabilized by electronic reorganization and structural reconstruction in the ultrathin limit. This is the most important point in this approach. To find the deviation from the $N^{-1}$ law, it is useful to calculate the following quantity: 
\begin{eqnarray}
p_\alpha(N) = \frac{d \ln \Delta E_\alpha }{d\ln N}. 
\label{eq:pN}
\end{eqnarray}
If $p_{\alpha}(N)=-1$, the $N$-layer thin film exactly follows the $N^{-1}$ law. The inequality of $p_{\alpha}(1)>-1$ indicates a downward deviation from the $N^{-1}$ law. 

A systematic calculation of FTEE has been performed for elements in the periodic table. As shown in Fig.~\ref{fig2}(a), a downward deviation from the $N^{-1}$ law is clearly observed for group IV elements (Si, Ge, and Sn). This is consistent with the experimental synthesis of silicene \cite{Vogt2012,Fleurence2012}, germanene \cite{Li2014,Bampoulis_2014,Davila_2014}, and stanene \cite{Zhu2015}. The values of $p(1)$ for elements in the periodic table are shown in Fig.~\ref{fig2}(b), predicting which elements prefer 2D structures. A similar deviation from the $N^{-1}$ law has been observed in III-V semiconductors \cite{SO2025}. In this way, studying the breakdown of the $N^{-1}$ law is useful for identifying non-vdW 2D materials.

The breakdown of the $N^{-1}$ scaling implies an electronic reorganization in the monolayer limit. For covalently bonded systems of silicene and germanene, this is attributed to the $sp^3$-to-$sp^2$ transformation. For metallic goldene, this is due to the electronic anisotropy around the Fermi level. As shown in Fig.~\ref{fig2}(c) and (d), the Fermi surface exhibits a roughly hexagonal shape, and the electron states on each edge have strong in-plane character and elongate along the edge normal direction, which forms the triangular bonding network in real space. The electronic anisotropy stabilizes the planar geometry. In contrast, aluminene (Al) exhibits a circular Fermi surface that consists of $p_x$, $p_y$, and $p_z$ states (see Fig.~\ref{fig2}(d)). The hybridization between in-plane and out-of-plane components may destabilize the planar geometry. To the best of our knowledge, a realization of aluminene has not been reported so far. 

\section{Summary}
In this perspective, we have reviewed the family of 2D materials including vdW and  non-vdW 2D materials. Several computational approaches to predict these systems have been explained. Geometrical screening-based approaches have identified layered structures in 3D materials, and a few thousand vdW 2D materials are predicted to be exfoliable. The elastic and force constants have been utilized in the screening of 3D materials databases, where many 2D materials different from those found by the geometry-based screening were identified. Still, it remains unclear how many candidate vdW 2D materials can be really synthesized experimentally.

High-throughput screening strategies for finding non-vdW 2D materials have yet to be established. Therefore, once a new 2D material is synthesized, a prototype-based approach has been used for the exploration. To resolve this, the author has highlighted the importance of the excess energy evolution in $N$ layered thin films, which enables us to find a 3D-2D transition, and proposed that the downward deviation from the $N^{-1}$ scaling should be a useful indicator for the non-vdW 2D materials. Application to elemental systems successfully predicted the experimentally synthesized 2D materials including silicene, germanene, and goldene. This partly explains the synthesizability of non-vdW 2D materials and provides a unified framework for understanding the emergent 2D structures beyond geometrical classification. 

It is physically transparent to rephrase vdW and non-vdW 2D materials as intrinsic and extrinsic 2D materials, respectively. The intrinsic 2D materials (e.g., graphene) are stable on their own and serve as a building block for constructing 3D bulk. On the other hand, the extrinsic 2D materials (e.g., silicene and goldene) become stable in the ultrathin limit, through electron redistribution and geometric relaxation. In other words, such parent systems possess an ``electronic flexibility'' to adopt the low-dimensional geometries with high surface-volume ratio. The two-dimensionality will vanish in the bulk limit because they are intrinsically 3D.  

Finally, although the substrate support is important in creating 2D structures, this perspective has focused on the intrinsic stability of free-standing 2D materials, which provides a starting point for more realistic simulation.


\begin{acknowledgments}
This work was supported by JSPS KAKENHI (Grant No. 24K01142). 
\end{acknowledgments}


\bibliography{refs}

@article{Novoselov2012,
	title = {A roadmap for graphene},
	volume = {490},
	issn = {1476-4687},
	url = {https://doi.org/10.1038/nature11458},
	doi = {10.1038/nature11458},
	number = {7419},
	journal = {Nature},
	author = {Novoselov, K. S. and Fal'ko, V. I. and Colombo, L. and Gellert, P. R. and Schwab, M. G. and Kim, K.},
	month = oct,
	year = {2012},
	pages = {192--200},
}

@article{Zhao2021,
author = {Zhao, Bei and Shen, Dingyi and Zhang, Zucheng and Lu, Ping and Hossain, Mongur and Li, Jia and Li, Bo and Duan, Xidong},
title = {{2D} Metallic Transition-Metal Dichalcogenides: Structures, Synthesis, Properties, and Applications},
journal = {Advanced Functional Materials},
volume = {31},
number = {48},
pages = {2105132},
doi = {https://doi.org/10.1002/adfm.202105132},
url = {https://advanced.onlinelibrary.wiley.com/doi/abs/10.1002/adfm.202105132},
year = {2021}
}

@article{Cahangirov2009,
  title = {Two- and One-Dimensional Honeycomb Structures of Silicon and Germanium},
  author = {Cahangirov, S. and Topsakal, M. and Akt\"urk, E. and \ifmmode \mbox{\c{S}}\else \c{S}\fi{}ahin, H. and Ciraci, S.},
  journal = {Phys. Rev. Lett.},
  volume = {102},
  issue = {23},
  pages = {236804},
  numpages = {4},
  year = {2009},
  month = {Jun},
  publisher = {American Physical Society},
  doi = {10.1103/PhysRevLett.102.236804},
  url = {https://link.aps.org/doi/10.1103/PhysRevLett.102.236804}
}

@article{Vogt2012,
  title = {Silicene: Compelling Experimental Evidence for Graphenelike Two-Dimensional Silicon},
  author = {Vogt, Patrick and De Padova, Paola and Quaresima, Claudio and Avila, Jose and Frantzeskakis, Emmanouil and Asensio, Maria Carmen and Resta, Andrea and Ealet, B\'en\'edicte and Le Lay, Guy},
  journal = {Phys. Rev. Lett.},
  volume = {108},
  issue = {15},
  pages = {155501},
  numpages = {5},
  year = {2012},
  month = {Apr},
  publisher = {American Physical Society},
  doi = {10.1103/PhysRevLett.108.155501},
  url = {https://link.aps.org/doi/10.1103/PhysRevLett.108.155501}
}

@article{Fleurence2012,
  title = {Experimental Evidence for Epitaxial Silicene on Diboride Thin Films},
  author = {Fleurence, Antoine and Friedlein, Rainer and Ozaki, Taisuke and Kawai, Hiroyuki and Wang, Ying and Yamada-Takamura, Yukiko},
  journal = {Phys. Rev. Lett.},
  volume = {108},
  issue = {24},
  pages = {245501},
  numpages = {5},
  year = {2012},
  month = {Jun},
  publisher = {American Physical Society},
  doi = {10.1103/PhysRevLett.108.245501},
  url = {https://link.aps.org/doi/10.1103/PhysRevLett.108.245501}
}

@article{Li2014,
author = {Li, Linfei and Lu, Shuang-Zan and Pan, Jinbo and Qin, Zhihui and Wang, Yu-Qi and Wang, Yeliang and Cao, Geng-Yu and Du, Shixuan and Gao, Hong-Jun},
title = {Buckled Germanene Formation on {Pt}(111)},
journal = {Advanced Materials},
volume = {26},
number = {28},
pages = {4820-4824},
doi = {https://doi.org/10.1002/adma.201400909},
url = {https://advanced.onlinelibrary.wiley.com/doi/abs/10.1002/adma.201400909},
year = {2014}
}

@article{Bampoulis_2014,
doi = {10.1088/0953-8984/26/44/442001},
url = {https://doi.org/10.1088/0953-8984/26/44/442001},
year = {2014},
month = {sep},
publisher = {IOP Publishing},
volume = {26},
number = {44},
pages = {442001},
author = {Bampoulis, P and Zhang, L and Safaei, A and van Gastel, R and Poelsema, B and Zandvliet, H J W},
title = {Germanene termination of {Ge}$_2${Pt} crystals on {Ge}(110)},
journal = {Journal of Physics: Condensed Matter}
}

@article{Davila_2014,
doi = {10.1088/1367-2630/16/9/095002},
url = {https://doi.org/10.1088/1367-2630/16/9/095002},
year = {2014},
month = {sep},
publisher = {IOP Publishing},
volume = {16},
number = {9},
pages = {095002},
author = {D{\'a}vila, M E and Xian, L and Cahangirov, S and Rubio, A and Le Lay, G},
title = {Germanene: a novel two-dimensional germanium allotrope akin to graphene and silicene},
journal = {New Journal of Physics}
}

@article{Zhu2015,
	title = {Epitaxial growth of two-dimensional stanene},
	volume = {14},
	issn = {1476-4660},
	url = {https://doi.org/10.1038/nmat4384},
	doi = {10.1038/nmat4384},
	number = {10},
	journal = {Nature Materials},
	author = {Zhu, Feng-Feng and Chen, Wei-Jiong and Xu, Yong and Gao, Chun-Lei and Guan, Dan-Dan and Liu, Can-Hua and Qian, Dong and Zhang, Shou-Cheng and Jia, Jin-Feng},
	month = oct,
	year = {2015},
	pages = {1020--1025},
}

@article{Yuhara2019,
author = {Yuhara, Junji and He, Bangjie and Matsunami, Noriaki and Nakatake, Masashi and Le Lay, Guy},
title = {Graphene's Latest Cousin: Plumbene Epitaxial Growth on a “Nano WaterCube”},
journal = {Advanced Materials},
volume = {31},
number = {27},
pages = {1901017},
doi = {https://doi.org/10.1002/adma.201901017},
url = {https://advanced.onlinelibrary.wiley.com/doi/abs/10.1002/adma.201901017},
year = {2019}
}

@article{Chung2024,
	title = {Structure and exfoliation mechanism of two-dimensional boron nanosheets},
	volume = {15},
	issn = {2041-1723},
	url = {https://doi.org/10.1038/s41467-024-49974-8},
	doi = {10.1038/s41467-024-49974-8},
	number = {1},
	journal = {Nature Communications},
	author = {Chung, Jing-Yang and Yuan, Yanwen and Mishra, Tara P. and Joseph, Chithralekha and Canepa, Pieremanuele and Ranjan, Pranay and Sadki, El Hadi S. and Grade{\v c}ak, Silvija and Garaj, Slaven},
	month = jul,
	year = {2024},
	pages = {6122}
}

@article{Kashiwaya2024,
	title = {Synthesis of goldene comprising single-atom layer gold},
	volume = {3},
	issn = {2731-0582},
	url = {https://doi.org/10.1038/s44160-024-00518-4},
	doi = {10.1038/s44160-024-00518-4},
	number = {6},
	journal = {Nature Synthesis},
	author = {Kashiwaya, Shun and Shi, Yuchen and Lu, Jun and Sangiovanni, Davide G. and Greczynski, Grzegorz and Magnuson, Martin and Andersson, Mike and Rosen, Johanna and Hultman, Lars},
	month = jun,
	year = {2024},
	pages = {744--751},
}

@Article{Molle2025,
author ="Molle, Alessandro and Yuhara, Junji and Yamada-Takamura, Yukiko and Sofer, Zdenek",
title  ="Synthesis of {Xenes}: physical and chemical methods",
journal  ="Chem. Soc. Rev.",
year  ="2025",
volume  ="54",
issue  ="4",
pages  ="1845-1869",
publisher  ="The Royal Society of Chemistry",
doi  ="10.1039/D4CS00999A",
url  ="http://dx.doi.org/10.1039/D4CS00999A"
}

@article{Acun_2015,
doi = {10.1088/0953-8984/27/44/443002},
url = {https://doi.org/10.1088/0953-8984/27/44/443002},
year = {2015},
month = {oct},
publisher = {IOP Publishing},
volume = {27},
number = {44},
pages = {443002},
author = {Acun, A and Zhang, L and Bampoulis, P and Farmanbar, M and van Houselt, A and Rudenko, A N and Lingenfelder, M and Brocks, G and Poelsema, B and Katsnelson, M I and Zandvliet, H J W},
title = {Germanene: the germanium analogue of graphene},
journal = {Journal of Physics: Condensed Matter}
}

@article{Bampoulis2023,
  title = {Quantum Spin Hall States and Topological Phase Transition in Germanene},
  author = {Bampoulis, Pantelis and Castenmiller, Carolien and Klaassen, Dennis J. and van Mil, Jelle and Liu, Yichen and Liu, Cheng-Cheng and Yao, Yugui and Ezawa, Motohiko and Rudenko, Alexander N. and Zandvliet, Harold J. W.},
  journal = {Phys. Rev. Lett.},
  volume = {130},
  issue = {19},
  pages = {196401},
  numpages = {6},
  year = {2023},
  month = {May},
  publisher = {American Physical Society},
  doi = {10.1103/PhysRevLett.130.196401},
  url = {https://link.aps.org/doi/10.1103/PhysRevLett.130.196401}
}

@article{Zhao2024,
  title = {Electrical conductivity of goldene},
  author = {Zhao, Shuo and Zhang, Huiwen and Zhu, Mingfeng and Jiang, Liwei and Zheng, Yisong},
  journal = {Phys. Rev. B},
  volume = {110},
  issue = {8},
  pages = {085111},
  numpages = {10},
  year = {2024},
  month = {Aug},
  publisher = {American Physical Society},
  doi = {10.1103/PhysRevB.110.085111},
  url = {https://link.aps.org/doi/10.1103/PhysRevB.110.085111}
}

@article{Malyi2019,
author = {Malyi, Oleksandr I. and Sopiha, Kostiantyn V. and Persson, Clas},
title = {Energy, Phonon, and Dynamic Stability Criteria of Two-Dimensional Materials},
journal = {ACS Applied Materials \& Interfaces},
volume = {11},
number = {28},
pages = {24876-24884},
year = {2019},
doi = {10.1021/acsami.9b01261},
URL = { 
        https://doi.org/10.1021/acsami.9b01261}
}

@article{lin_general_2022,
	title = {General invariance and equilibrium conditions for lattice dynamics in {1D}, {2D}, and {3D} materials},
	volume = {8},
	issn = {2057-3960},
	url = {https://doi.org/10.1038/s41524-022-00920-6},
	doi = {10.1038/s41524-022-00920-6},
	number = {1},
	journal = {npj Computational Materials},
	author = {Lin, Changpeng and Ponc{\'e}, Samuel and Marzari, Nicola},
	month = nov,
	year = {2022},
	pages = {236},
}

@article{MP2013,
	title = {Commentary: {The} {Materials} {Project}: {A} materials genome approach to accelerating materials innovation},
	volume = {1},
	issn = {2166-532X},
	url = {https://doi.org/10.1063/1.4812323},
	doi = {10.1063/1.4812323},
	number = {1},
	journal = {APL Materials},
	author = {Jain, Anubhav and Ong, Shyue Ping and Hautier, Geoffroy and Chen, Wei and Richards, William Davidson and Dacek, Stephen and Cholia, Shreyas and Gunter, Dan and Skinner, David and Ceder, Gerbrand and Persson, Kristin A.},
	month = jul,
	year = {2013},
	pages = {011002},
}

@article{Larsen2019,
  title = {Definition of a scoring parameter to identify low-dimensional materials components},
  author = {Larsen, Peter Mahler and Pandey, Mohnish and Strange, Mikkel and Jacobsen, Karsten Wedel},
  journal = {Phys. Rev. Mater.},
  volume = {3},
  issue = {3},
  pages = {034003},
  numpages = {11},
  year = {2019},
  month = {Mar},
  publisher = {American Physical Society},
  doi = {10.1103/PhysRevMaterials.3.034003},
  url = {https://link.aps.org/doi/10.1103/PhysRevMaterials.3.034003}
}

@article{ASE,
doi = {10.1088/1361-648X/aa680e},
url = {https://doi.org/10.1088/1361-648X/aa680e},
year = {2017},
month = {jun},
publisher = {IOP Publishing},
volume = {29},
number = {27},
pages = {273002},
author = {Hjorth Larsen, Ask and J{\o}rgen Mortensen, Jens and Blomqvist, Jakob and
Castelli, Ivano E and Christensen, Rune and Du{\l}ak, Marcin and Friis, Jesper
and Groves, Michael N and Hammer, Bj{\o}rk and Hargus, Cory and Hermes, Eric D
and Jennings, Paul C and Bjerre Jensen, Peter and Kermode, James and Kitchin,
John R and Leonhard Kolsbjerg, Esben and Kubal, Joseph and Kaasbjerg, Kristen
and Lysgaard, Steen and Bergmann Maronsson, J{'o}n and Maxson, Tristan and
Olsen, Thomas and Pastewka, Lars and Peterson, Andrew and Rostgaard, Carsten
and Schi{\o}tz, Jakob and Sch{\"u}tt, Ole and Strange, Mikkel and Thygesen,
Kristian S and Vegge, Tejs and Vilhelmsen, Lasse and Walter, Michael and
Zeng, Zhenhua and Jacobsen, Karsten W},
title = {The atomic simulation environment—a Python library for working with atoms},
journal = {Journal of Physics: Condensed Matter}
}

@article{Jung2018,
author = {Jung, Jong Hyun and Park, Cheol-Hwan and Ihm, Jisoon},
title = {A Rigorous Method of Calculating Exfoliation Energies from First Principles},
journal = {Nano Letters},
volume = {18},
number = {5},
pages = {2759-2765},
year = {2018},
doi = {10.1021/acs.nanolett.7b04201},
URL = {    
        https://doi.org/10.1021/acs.nanolett.7b04201
}}

@article{Lebegue2013,
  title = {Two-Dimensional Materials from Data Filtering and Ab Initio Calculations},
  author = {Leb\`egue, S. and Bj\"orkman, T. and Klintenberg, M. and Nieminen, R. M. and Eriksson, O.},
  journal = {Phys. Rev. X},
  volume = {3},
  issue = {3},
  pages = {031002},
  numpages = {7},
  year = {2013},
  month = {Jul},
  publisher = {American Physical Society},
  doi = {10.1103/PhysRevX.3.031002},
  url = {https://link.aps.org/doi/10.1103/PhysRevX.3.031002}
}

@article{Ashton2017,
  title = {Topology-Scaling Identification of Layered Solids and Stable Exfoliated 2D Materials},
  author = {Ashton, Michael and Paul, Joshua and Sinnott, Susan B. and Hennig, Richard G.},
  journal = {Phys. Rev. Lett.},
  volume = {118},
  issue = {10},
  pages = {106101},
  numpages = {6},
  year = {2017},
  month = {Mar},
  publisher = {American Physical Society},
  doi = {10.1103/PhysRevLett.118.106101},
  url = {https://link.aps.org/doi/10.1103/PhysRevLett.118.106101}
}

@article{Mounet2018,
	title = {Two-dimensional materials from high-throughput computational exfoliation of experimentally known compounds},
	volume = {13},
	issn = {1748-3395},
	url = {https://doi.org/10.1038/s41565-017-0035-5},
	doi = {10.1038/s41565-017-0035-5},
	number = {3},
	journal = {Nature Nanotechnology},
	author = {Mounet, Nicolas and Gibertini, Marco and Schwaller, Philippe and Campi, Davide and Merkys, Andrius and Marrazzo, Antimo and Sohier, Thibault and Castelli, Ivano Eligio and Cepellotti, Andrea and Pizzi, Giovanni and Marzari, Nicola},
	month = mar,
	year = {2018},
	pages = {246--252},
}

@article{Mounet2023,
author = {Campi, Davide and Mounet, Nicolas and Gibertini, Marco and Pizzi, Giovanni and Marzari, Nicola},
title = {Expansion of the Materials Cloud {2D} Database},
journal = {ACS Nano},
volume = {17},
number = {12},
pages = {11268-11278},
year = {2023},
doi = {10.1021/acsnano.2c11510},
URL = { 
        https://doi.org/10.1021/acsnano.2c11510
}
}

@article{bonilla_2018,
	title = {Strong room-temperature ferromagnetism in {VSe}$_2$ monolayers on van der {Waals} substrates},
	volume = {13},
	issn = {1748-3395},
	url = {https://doi.org/10.1038/s41565-018-0063-9},
	doi = {10.1038/s41565-018-0063-9},
	number = {4},
	journal = {Nature Nanotechnology},
	author = {Bonilla, Manuel and Kolekar, Sadhu and Ma, Yujing and Diaz, Horacio Coy and Kalappattil, Vijaysankar and Das, Raja and Eggers, Tatiana and Gutierrez, Humberto R. and Phan, Manh-Huong and Batzill, Matthias},
	month = apr,
	year = {2018},
	pages = {289--293},
}

@article{Haastrup_2018,
doi = {10.1088/2053-1583/aacfc1},
url = {https://doi.org/10.1088/2053-1583/aacfc1},
year = {2018},
month = {sep},
publisher = {IOP Publishing},
volume = {5},
number = {4},
pages = {042002},
author = {Haastrup, Sten and Strange, Mikkel and Pandey, Mohnish and Deilmann, Thorsten and Schmidt, Per S and Hinsche, Nicki F and Gjerding, Morten N and Torelli, Daniele and Larsen, Peter M and Riis-Jensen, Anders C and Gath, Jakob and Jacobsen, Karsten W and J{\o}rgen Mortensen, Jens and Olsen, Thomas and Thygesen, Kristian S},
title = {The Computational {2D} Materials Database: high-throughput modeling and discovery of atomically thin crystals},
journal = {2D Materials}
}

@article{Gjerding_2021,
doi = {10.1088/2053-1583/ac1059},
url = {https://doi.org/10.1088/2053-1583/ac1059},
year = {2021},
month = {jul},
publisher = {IOP Publishing},
volume = {8},
number = {4},
pages = {044002},
author = {Gjerding, Morten Niklas and Taghizadeh, Alireza and Rasmussen, Asbj{\o}rn and Ali, Sajid and Bertoldo, Fabian and Deilmann, Thorsten and Kn{\o}sgaard, Nikolaj R\UTF{00F8}rb\UTF{00E6}k and Kruse, Mads and Larsen, Ask Hjorth and Manti, Simone and Pedersen, Thomas Garm and Petralanda, Urko and Skovhus, Thorbj{\o}rn and Svendsen, Mark Kamper and Mortensen, Jens J{\o}rgen and Olsen, Thomas and Thygesen, Kristian Sommer},
title = {Recent progress of the Computational {2D} Materials Database ({C2DB})},
journal = {2D Materials}
}

@article{lyngby_2022,
	title = {Data-driven discovery of {2D} materials by deep generative models},
	volume = {8},
	issn = {2057-3960},
	url = {https://doi.org/10.1038/s41524-022-00923-3},
	doi = {10.1038/s41524-022-00923-3},
	number = {1},
	journal = {npj Computational Materials},
	author = {Lyngby, Peder and Thygesen, Kristian Sommer},
	month = nov,
	year = {2022},
	pages = {232},
}

@article{Wang_2023,
doi = {10.1088/2053-1583/accc43},
url = {https://doi.org/10.1088/2053-1583/accc43},
year = {2023},
month = {apr},
publisher = {IOP Publishing},
volume = {10},
number = {3},
pages = {035007},
author = {Wang, Hai-Chen and Schmidt, Jonathan and Marques, Miguel A L and Wirtz, Ludger and Romero, Aldo H},
title = {Symmetry-based computational search for novel binary and ternary {2D} materials},
journal = {2D Materials}
}

@article{X2DB,
doi = {10.48550/arXiv.2603.05083},
url = {https://doi.org/10.48550/arXiv.2603.05083},
author = {Mohammad A. Akhound and Tara M. Boland and Mikkel O. Sauer and Matthias Batzill and Moses A. Bokinala and Stela Canulescu and Yury Gogotsi and Philip Hofmann and Andras Kis and Jiong Lu and Thomas Michely and S{\o}ren Raza and Wencai Ren and Joshua A. Robinson and Zdenek Sofer and Jing H. Teng and S{\o}ren Ulstrup and Meng Zhao and Xiaoxu Zhao and Jens J. Mortensen and Thomas Olsen and Kristian S. Thygesen},
title = {Large-scale Integration of Experimental and Computational Data for {2D} Materials},
journal = {arXiv}
}

@article{Hong2020,
author = {Yi-Lun Hong  and Zhibo Liu  and Lei Wang  and Tianya Zhou  and Wei Ma  and Chuan Xu  and Shun Feng  and Long Chen  and Mao-Lin Chen  and Dong-Ming Sun  and Xing-Qiu Chen  and Hui-Ming Cheng  and Wencai Ren },
title = {Chemical vapor deposition of layered two-dimensional {MoSi}$_2${N}$_4$ materials},
journal = {Science},
volume = {369},
number = {6504},
pages = {670-674},
year = {2020},
doi = {10.1126/science.abb7023},
URL = {https://www.science.org/doi/abs/10.1126/science.abb7023},
}

@article{wang_intercalated_2021,
	title = {Intercalated architecture of {MA$_2$Z$_4$} family layered van der {Waals} materials with emerging topological, magnetic and superconducting properties},
	volume = {12},
	issn = {2041-1723},
	url = {https://doi.org/10.1038/s41467-021-22324-8},
	doi = {10.1038/s41467-021-22324-8},
	number = {1},
	journal = {Nature Communications},
	author = {Wang, Lei and Shi, Yongpeng and Liu, Mingfeng and Zhang, Ao and Hong, Yi-Lun and Li, Ronghan and Gao, Qiang and Chen, Mingxing and Ren, Wencai and Cheng, Hui-Ming and Li, Yiyi and Chen, Xing-Qiu},
	month = apr,
	year = {2021},
	pages = {2361},
}

@article{Jia2021,
	title = {Elasticity-based-exfoliability measure for high-throughput computational exfoliation of two-dimensional materials},
	volume = {7},
	issn = {2057-3960},
	url = {https://doi.org/10.1038/s41524-021-00677-4},
	doi = {10.1038/s41524-021-00677-4},
	number = {1},
	journal = {npj Computational Materials},
	author = {Jia, Xiangzheng and Shao, Qian and Xu, Yongchun and Li, Ruishan and Huang, Kai and Guo, Yongzhe and Qu, Cangyu and Gao, Enlai},
	month = dec,
	year = {2021},
	pages = {211},
}

@article{Bagheri2023,
author = {Bagheri, Mohammad and Berger, Ethan and Komsa, Hannu-Pekka},
title = {Identification of Material Dimensionality Based on Force Constant Analysis},
journal = {The Journal of Physical Chemistry Letters},
volume = {14},
number = {35},
pages = {7840-7847},
year = {2023},
doi = {10.1021/acs.jpclett.3c01635},
URL = { 
        https://doi.org/10.1021/acs.jpclett.3c01635},
}

@article{TOGO20151,
title = {First principles phonon calculations in materials science},
journal = {Scripta Materialia},
volume = {108},
pages = {1-5},
year = {2015},
issn = {1359-6462},
doi = {https://doi.org/10.1016/j.scriptamat.2015.07.021},
url = {https://www.sciencedirect.com/science/article/pii/S1359646215003127},
author = {Atsushi Togo and Isao Tanaka}
}

@article{Bagheri2026,
author = {Bagheri, Mohammad and Berger, Ethan and Komsa, Hannu-Pekka and Koskinen, Pekka},
title = {Massive Discovery of Low-Dimensional Materials from Universal Computational Strategy},
journal = {Chemistry of Materials},
volume = {38},
number = {5},
pages = {2395-2402},
year = {2026},
doi = {10.1021/acs.chemmater.5c03151},
URL = {https://doi.org/10.1021/acs.chemmater.5c03151}
}

@article{Lucking2018,
  title = {Traditional Semiconductors in the Two-Dimensional Limit},
  author = {Lucking, Michael C. and Xie, Weiyu and Choe, Duk-Hyun and West, Damien and Lu, Toh-Ming and Zhang, S. B.},
  journal = {Phys. Rev. Lett.},
  volume = {120},
  issue = {8},
  pages = {086101},
  numpages = {5},
  year = {2018},
  month = {Feb},
  publisher = {American Physical Society},
  doi = {10.1103/PhysRevLett.120.086101},
  url = {https://link.aps.org/doi/10.1103/PhysRevLett.120.086101}
}

@article{Zheng_2015,
  title = {Monolayer II-VI semiconductors: A first-principles prediction},
  author = {Zheng, Hui and Li, Xian-Bin and Chen, Nian-Ke and Xie, Sheng-Yi and Tian, Wei Quan and Chen, Yuanping and Xia, Hong and Zhang, S. B. and Sun, Hong-Bo},
  journal = {Phys. Rev. B},
  volume = {92},
  issue = {11},
  pages = {115307},
  numpages = {10},
  year = {2015},
  month = {Sep},
  publisher = {American Physical Society},
  doi = {10.1103/PhysRevB.92.115307},
  url = {https://link.aps.org/doi/10.1103/PhysRevB.92.115307}
}

@article{SO2022,
author = {Ono ,Shota},
title = {Two-Dimensional Ionic Crystals: The Cases of {IA}-{VII} Alkali Halides and {IA}-{IB} CsAu},
journal = {Journal of the Physical Society of Japan},
volume = {91},
number = {9},
pages = {094606},
year = {2022},
doi = {10.7566/JPSJ.91.094606},
URL = {https://doi.org/10.7566/JPSJ.91.094606}
}

@article{Qin_2021,
author = {Qin, Le and Zhang, Zhi-Hao and Jiang, Zeyu and Fan, Kai and Zhang, Wen-Hao and Tang, Qiao-Yin and Xia, Hui-Nan and Meng, Fanqi and Zhang, Qinghua and Gu, Lin and West, Damien and Zhang, Shengbai and Fu, Ying-Shuang},
title = {Realization of {AlSb} in the Double-Layer Honeycomb Structure: A Robust Class of Two-Dimensional Material},
journal = {ACS Nano},
volume = {15},
number = {5},
pages = {8184-8191},
year = {2021},
doi = {10.1021/acsnano.1c00470},
URL = {https://doi.org/10.1021/acsnano.1c00470}
}

@article{Balan_2018_hematite,
	title = {Exfoliation of a non-van der {Waals} material from iron ore hematite},
	volume = {13},
	issn = {1748-3395},
	url = {https://doi.org/10.1038/s41565-018-0134-y},
	doi = {10.1038/s41565-018-0134-y},
	number = {7},
	journal = {Nature Nanotechnology},
	author = {Puthirath Balan, Aravind and Radhakrishnan, Sruthi and Woellner, Cristiano F. and Sinha, Shyam K. and Deng, Liangzi and Reyes, Carlos de los and Rao, Banki Manmadha and Paulose, Maggie and Neupane, Ram and Apte, Amey and Kochat, Vidya and Vajtai, Robert and Harutyunyan, Avetik R. and Chu, Ching-Wu and Costin, Gelu and Galvao, Douglas S. and Mart{\' i}, Angel A. and van Aken, Peter A. and Varghese, Oomman K. and Tiwary, Chandra Sekhar and Malie Madom Ramaswamy Iyer, Anantharaman and Ajayan, Pulickel M.},
	month = jul,
	year = {2018},
	pages = {602--609},
}

@article{Balan_2018_ilmenite,
author = {Puthirath Balan, Aravind and Radhakrishnan, Sruthi and Kumar, Ritesh and Neupane, Ram and Sinha, Shyam Kanta and Deng, Liangzi and de los Reyes, Carlos A. and Apte, Amey and Rao, B. Manmadha and Paulose, Maggie and Vajtai, Robert and Chu, Ching Wu and Costin, Gelu and Mart{\' i}, Angel A. and Varghese, Oomman K. and Singh, Abhishek K. and Tiwary, Chandra Sekhar and Anantharaman, Maliemadom R. and Ajayan, Pulickel M.},
title = {A Non-van der Waals Two-Dimensional Material from Natural Titanium Mineral Ore Ilmenite},
journal = {Chemistry of Materials},
volume = {30},
number = {17},
pages = {5923-5931},
year = {2018},
doi = {10.1021/acs.chemmater.8b01935},
URL = { 
        https://doi.org/10.1021/acs.chemmater.8b01935
}
}

@article{Friedrich2022,
author = {Friedrich, Rico and Ghorbani-Asl, Mahdi and Curtarolo, Stefano and Krasheninnikov, Arkady V.},
title = {Data-Driven Quest for Two-Dimensional Non-van der Waals Materials},
journal = {Nano Letters},
volume = {22},
number = {3},
pages = {989-997},
year = {2022},
doi = {10.1021/acs.nanolett.1c03841},
URL = {     
        https://doi.org/10.1021/acs.nanolett.1c03841
}
}

@article{Nevalaita2018,
  title = {Atlas for the properties of elemental two-dimensional metals},
  author = {Nevalaita, Janne and Koskinen, Pekka},
  journal = {Phys. Rev. B},
  volume = {97},
  issue = {3},
  pages = {035411},
  numpages = {11},
  year = {2018},
  month = {Jan},
  publisher = {American Physical Society},
  doi = {10.1103/PhysRevB.97.035411},
  url = {https://link.aps.org/doi/10.1103/PhysRevB.97.035411}
}

@article{SO2020,
  title = {Dynamical stability of two-dimensional metals in the periodic table},
  author = {Ono, Shota},
  journal = {Phys. Rev. B},
  volume = {102},
  issue = {16},
  pages = {165424},
  numpages = {14},
  year = {2020},
  month = {Oct},
  publisher = {American Physical Society},
  doi = {10.1103/PhysRevB.102.165424},
  url = {https://link.aps.org/doi/10.1103/PhysRevB.102.165424}
}

@article{SO2025,
  title = {Framework for identifying non--van der Waals two-dimensional materials},
  author = {Ono, Shota},
  journal = {Phys. Rev. B},
  volume = {112},
  issue = {7},
  pages = {075403},
  numpages = {11},
  year = {2025},
  month = {Aug},
  publisher = {American Physical Society},
  doi = {10.1103/crw6-zvpx},
  url = {https://link.aps.org/doi/10.1103/crw6-zvpx}
}

@article{SO_HY_2025,
  title = {Anomalous chirality dependence of strain energy in gold nanotubes},
  author = {Ono, Shota and Yoshioka, Hideo},
  journal = {Phys. Rev. B},
  volume = {111},
  issue = {8},
  pages = {085414},
  numpages = {6},
  year = {2025},
  month = {Feb},
  publisher = {American Physical Society},
  doi = {10.1103/PhysRevB.111.085414},
  url = {https://link.aps.org/doi/10.1103/PhysRevB.111.085414}
}

@article{qe,
	doi = {10.1088/1361-648x/aa8f79},
	url = {https://doi.org/10.1088%2F1361-648x%2Faa8f79},
	year = 2017,
	month = {oct},
	publisher = {{IOP} Publishing},
	volume = {29},
	number = {46},
	pages = {465901},
	author = {P Giannozzi and O Andreussi and T Brumme and O Bunau and M Buongiorno Nardelli and M Calandra and R Car and C Cavazzoni and D Ceresoli and M Cococcioni and N Colonna and I Carnimeo and A Dal Corso and S de Gironcoli and P Delugas and R A DiStasio and A Ferretti and A Floris and G Fratesi and G Fugallo and R Gebauer and U Gerstmann and F Giustino and T Gorni and J Jia and M Kawamura and H-Y Ko and A Kokalj and E K{\"u}{\c{c}}{\"u}kbenli and M Lazzeri and M Marsili and N Marzari and F Mauri and N L Nguyen and H-V Nguyen and A Otero-de-la-Roza and L Paulatto and S Ponc{\'{e}} and D Rocca and R Sabatini and B Santra and M Schlipf and A P Seitsonen and A Smogunov and I Timrov and T Thonhauser and P Umari and N Vast and X Wu and S Baroni},
	title = {Advanced capabilities for materials modelling with Quantum {ESPRESSO}},
	journal = {Journal of Physics: Condensed Matter}
}

@article{vesta,
  title = {VESTA 3 for three-dimensional visualization of crystal, volumetric and morphology data},
  author = {Momma, K and Izumi, F},
  journal = {J. Appl. Crystallogr.},
  volume = {44},
  pages = {1272},
  year = {2011},
  url = {https://doi.org/10.1107/S0021889811038970}
}

\end{document}